\documentclass[a4,12pt]{article}
\usepackage{graphicx,psfrag,amssymb,cite}
\renewcommand{\thefootnote}{\fnsymbol{footnote}}

\newcommand{\prepr}[1] {\begin{flushright}  {\bf #1} \end{flushright} \vskip 1.cm}
\newcommand{\titul}[1] {\boldmath \begin{center}{\Large {\bf #1 } } \end{center}
\vskip 0.8cm}

\newcommand{\autor}[1] {\begin{center}  {\bf \lineskip .3cm #1  }
                        \end{center} }

\newcommand{\lugar}[1] {\begin{center}  {\normalsize \bf \it #1   } \end{center}}
\newcounter{muni}

\makeatletter
\def\fmslash{\@ifnextchar[{\fmsl@sh}{\fmsl@sh[0mu]}}
\def\fmsl@sh[#1]#2{%
  \mathchoice
    {\@fmsl@sh\displaystyle{#1}{#2}}%
    {\@fmsl@sh\textstyle{#1}{#2}}%
    {\@fmsl@sh\scriptstyle{#1}{#2}}%
    {\@fmsl@sh\scriptscriptstyle{#1}{#2}}}
\def\@fmsl@sh#1#2#3{\m@th\ooalign{$\hfil#1\mkern#2/\hfil$\crcr$#1#3$}}
\makeatother
\begin{document}
\hbadness=10000
\pagenumbering{arabic}
\begin{titlepage}

\prepr{hep-ph/0210376\\
\hspace{30mm} KIAS--P02071 \\
\hspace{30mm} KEK--TH--852 \\
\hspace{30mm} October 2002}

\begin{center}
\titul{\bf Large enhancement of $D^\pm\to e^\pm\nu$
and $D^\pm_s \to e^\pm\nu$ in\\ $R$ Parity violating SUSY models}

\autor{A.G. Akeroyd$^{\mbox{1}}$\footnote{akeroyd@kias.re.kr},
S. Recksiegel$^{\mbox{2}}$\footnote{stefan@post.kek.jp} }
\lugar{ $^{1}$ Korea Institute for Advanced Study,
207-43 Cheongryangri-dong,\\ Dongdaemun-gu,
Seoul 130-012, Republic of Korea}

\lugar{ $^{2}$ Theory Group, KEK, Tsukuba, Ibaraki 305-0801, Japan }

\end{center}

\vskip2.0cm

\begin{abstract}
\noindent{The purely leptonic decays $D^\pm\to e^\pm\nu$
and $D^\pm_s\to e^\pm\nu$, for which no experimental
limits exist, are highly suppressed in the 
Standard Model. Mere observation of these decays
at the $B$ factories BELLE/BaBar or forthcoming CLEO-c would be 
a clear signal of physics beyond the SM. We show that $R$ parity violating 
slepton contributions can give rise to 
spectacular enhancements of the decay
rates, resulting in branching ratios as large as $5\times 10^{-3}$, which 
strongly motivates a search in these channels.}

\end{abstract}

\vskip1.0cm
%{\bf  PACS index : }
\end{titlepage}
\thispagestyle{empty}
\newpage

\pagestyle{plain}
\renewcommand{\thefootnote}{\arabic{footnote} }
\setcounter{footnote}{0}

\section{Introduction}

The wealth of new data from the $B$ factories BELLE and BaBar has
caused a great amount of phenomenological interest in $B$ decays in 
recent years. Already the much anticipated measurement
of $\sin2\phi_1$ \cite{Abe:2001xe,Aubert:2001nu} has been 
achieved, and many new results in the field of rare $B$ decays 
(e.g. $b\to s\gamma$ \cite{Bertolini:1990if}
 and $b\to d\gamma$ \cite{Akeroyd:2001cy}) are eagerly awaited.

Less attention has been devoted to charmed ($D$) meson decays, 
although the $B$ factories and forthcoming CLEO--c 
promise the largest sample of charmed mesons to date. 
$D$ mesons may be produced at the $B$ factories by
two mechanisms: i) Continuum $c\overline c$ production
and ii) Decay of $B$ mesons.
While the majority of $B$ decays involve some charmed particles,
it is difficult to extract charm data from these decays due to the high
multiplicity of particles in the final states. With the high luminosity
of the $B$ factories, however, a lot of $c\bar c$ pairs that subsequently
hadronize to $D$ mesons are produced directly in the 
collision of the primary $e^+e^-$ beams. Both BELLE and BaBar will each
have about $5\times 10^{8}$ $c\overline c$ continuum events in the 
anticipated data samples of $400$ fb$^{-1}$, thus providing
a rich testing ground for charm decays. At CLEO-c, prospects are
also very promising with the threshold production of $D$ mesons 
offering distinct advantages over the $c\overline c$
continuum production at the $B$ factories, which compensates
for the lower luminosity of CLEO-c \cite{Shipsey:2002ye}.

These experiments will have substantially increased
sensitivity to the purely leptonic decays of the charged $D^\pm$ mesons, 
$D^\pm_s\to l^\pm\nu$ and $D^\pm\to l^\pm\nu$. 
In the SM, such decays occur via $W^\pm$ annihilation in the $s$-channel
and provide an opportunity to measure the decay constants ($f_{D_s},f_D$)
for $D^\pm_s$ and $D^\pm$. Of these six leptonic decays, only 
$D^\pm_s\to \tau^\pm\nu$ and $D^\pm_s\to \mu^\pm\nu$ have
been observed, from which $f_{D_s}$ is measured with an error $\sim 14\%$. 
The $D^\pm$ decays are Cabbibo suppressed compared to $D^\pm_s$, 
and none have been observed except for 1 event for $D^\pm\to \mu^\pm\nu$.
The $B$ factories and CLEO-c will offer improved measurements
of $D^\pm_s\to \tau^\pm\nu$ and $D^\pm_s\to \mu^\pm\nu$, which in turn 
will significantly reduce the error in the current measurements of 
$f_{D_s}$. Observation of $D^\pm\to \mu^\pm\nu$ (or the less accessible
$D^\pm\to \tau^\pm\nu$) will provide the first serious 
measurement of $f_{D}$. 

In this paper we advocate searching for physics beyond the SM 
through these decays. Of special interest are 
$D^\pm_s\to e^\pm\nu$ and $D^\pm\to e^\pm\nu$ 
for which no experimental limits exist, but could readily be 
searched for at the above experiments. In the SM 
these are severely helicity suppressed by $m_e^2$ and have BRs of 
order $10^{-7}$ and $10^{-9}$, respectively. Hence such decays
have been largely overlooked since (in the context of the SM) they
cannot offer a measurement of the decay constant with present
or upcoming data samples. However, the smallness of their BRs 
enables these decays to play a new role of probing models beyond the SM. 
We show that slepton contributions in the Minimal Supersymmetric 
Standard Model (MSSM) with explicit $R$ parity violation 
can enhance these BRs to $5\times 10^{-3}$, 
a result which strongly motivates a search in these channels.
Although the effect of new physics in purely leptonic decays is
sometimes tainted by the uncertainty in the decay constant, the
tiny SM branching ratios for $D^\pm_s\to e^\pm\nu$ and $D^\pm\to e^\pm\nu$ 
assure that mere observation of these decays at the aforementioned 
machines would be an unambiguous signal of physics beyond the SM.

Our work is organised as follows. In section 2 we introduce the
$D^\pm$ meson annihilation decays. In section 3 we show how these decays can
be enhanced in $R$ parity violating SUSY models. Section 4 presents our
numerical results and section 5 contains our conclusions. 

%%%%%%%%%%%%%%%%%%%%%%%%%%%%%%%%%%%%%%%%%%%%%%%%%%%%%%%%%%
\boldmath
\section{Annihilation $D^\pm$ meson decays.}
\unboldmath
%%%%%%%%%%%%%%%%%%%%%%%%%%%%%%%%%%%%%%%%%%%%%%%%%%%%%%%%%%

To date the primary interest in measuring the purely leptonic decays 
$D^\pm/D^\pm_s\to l^\pm\nu_l$
has been to obtain information on the charged $D$ meson decay constants 
\cite{Soldner-Rembold:2001zk}. 
In the SM these decays proceed via annihilation to a $W^\pm$ in the 
$s-$channel (see Fig.1). Due to helicity suppression, the rate is 
proportional to $m^2_l$, and the phase space suppression is 
particularly severe for $\tau^\pm\nu$. The BR is given by:
\begin{equation} \label{smrate}
\Gamma(D^+_q\to \ell^+\nu_\ell)={G_F^2 m_{D_q} m_l^2 f_{D_q}^2\over 8\pi}
|V_{cq}|^2 \left(1-{m_l^2\over m^2_{D_q}}\right)^2
\end{equation}

\begin{figure} \begin{center}
\includegraphics[width=7cm]{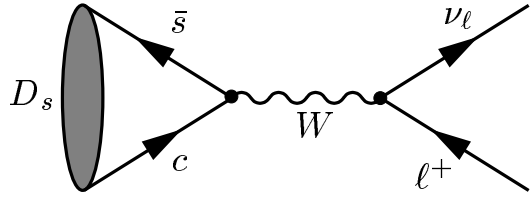}
\caption{Leptonic $D_s$ decay in the Standard Model: $W$ exchange
\label{wexchange}}
\end{center} \end{figure}

\begin{table}\begin{center}
\begin{tabular} {|c|c|c|c|c|} \hline
Decay & SM Prediction & Experiment \\ \hline
 $D_d^+\to e^+\nu_e$ & $8.24\times 10^{-9}$ & $\times$ \\ \hline
 $D_d^+\to \mu^+\nu_{\mu}$ & $3.50\times 10^{-4}$ & 
     $(8^{\,\,+16\,\,+5}_{\,\,\,\,-5\,\,\, -2}\,)
     \times 10^{-4}$ \cite{Bai:cg}\\ \hline
 $D_d^+\to \tau^+\nu_{\tau} $ & $9.25\times 10^{-4}$ & $\times$  \\ \hline
 $D_s^+\to e^+\nu_e$ & $1.23\times 10^{-7}$ & $\times$ \\ \hline
 $D_s^+\to \mu^+\nu_{\mu}$ & $5.22\times 10^{-3}$ & 
     $(5.3\pm 0.9\pm 1.2)\times 10^{-3}$ \cite{Soldner-Rembold:2001zk}\\ \hline
 $D_s^+\to \tau^+\nu_{\tau}$ & $5.09\times 10^{-2}$ & 
     $(6.05\pm 1.04\pm 1.34\pm 0.22)\times 10^{-2}$ \cite{Soldner-Rembold:2001zk} \\ \hline
\end{tabular}\end{center}
\caption{SM predictions and current experimental limits. \label{explimits}}
\end{table}
\noindent
where $q=d$ or $s$. The SM predictions for the BRs and the current 
experimental status of the various searches are shown in 
Table 1.\footnote{Our numbers differ substantially --- especially
in the $\tau$ channels --- from the ones given e.g.\ in the BaBar
book. This is because the rate depends very strongly on the mass
of the $D$, for which we are using newer values. We also use
a different value for the decay constant of the $D_s$, see section
4 for details.}
One can see that the decays involving $e^\pm\nu$ have tiny BRs, while
those involving $\mu^\pm\nu$ and $\tau^\pm\nu$ have BRs in the range
$10^{-2}\to 10^{-4}$. Of the six possible decays, only two have been 
measured with any sort of accuracy, yielding a world average of
$f_{D_s}=264\pm 35$ MeV \cite{Soldner-Rembold:2001zk}. 
Additionally there is a very imprecise 
measurement of $D^\pm\to \mu^\pm\nu$ based on 1 observed event,
giving $f_{D_s}=300^{+180+80}_{-150-40}$ MeV \cite{Bai:cg}. 
Of the three decays
which have not been searched for, $D^\pm_s\to e^\pm\nu$ and 
$D^\pm\to e^\pm\nu$ have particularly clean signatures. With the 
expected large samples of $D^\pm$ and $D^\pm_s$ mesons at BELLE, BaBar and 
CLEO-c \cite{Shipsey:2002ye}, these experiments 
should be sensitive to BR$\sim {\cal O} (10^{-4})$. CLEO-c aims to
accumulate 30 million $D^\pm$ events (6 million fully tagged)
and 1.5 million $D^\pm_s$ events (0.3 million fully tagged)
by the end of 2004. BELLE and BaBar expect $5\times 10^8$
$c\overline c$ continuum events by the end of 2005.
Mere observation of these decays would be an patent 
signal of physics beyond the SM. In the next section we show that 
SUSY particles
in $R$ parity violating extensions of the MSSM can enhance these decays to 
experimental observability. Thus in addition to offering measurements
of the charged $D$ meson decay constants, the purely leptonic decays of 
$D^\pm/D^\pm_s$ mesons assume a new role of probing physics beyond the SM.
In \cite{Shipsey:2002ye} CLEO-c is considering increasing the 
selection efficiency of the $\mu$--channel by waiving the 
$\mu$--tag requirement, stating
that no $e$--contamination is to be expected due to the small SM
rate of the respective channel. We strongly encourage also performing
an analysis with a muon identification tag, because the $e$--channel 
can substantially
contribute to the total leptonic annihilation decay rate as will be
shown later in this paper.

The related decays $D^\pm/D^\pm_s\to e^\pm\nu \gamma$, are known to have
larger BRs than $D^\pm/D^\pm_s\to e^\pm\nu$ in the SM
\cite{Burdman:1994ip,Korchemsky:1999qb}. This is because the
presence of a photon in the final state removes the helicity
suppression. The analysis of \cite{Korchemsky:1999qb} finds
BR$(D^\pm\to e^\pm\nu \gamma) \sim {\cal O} (10^{-4}\to 10^{-5})$
and BR$(D_s^\pm\to e^\pm\nu \gamma) \sim {\cal O} (10^{-3}\to 10^{-4}$)
The ``effective'' SM prediction for
BR$(D^\pm/D^\pm_s\to e^\pm\nu$) should include
the contribution from BR$(D^\pm/D^\pm_s\to e^\pm\nu \gamma$) 
with a soft photon (i.e. one which cannot be detected experimentally),
whose infra-red singularity cancels with the radiative 
corrections to BR($D^\pm/D^\pm_s\to e^\pm\nu$):
\begin{equation}
BR^{\rm eff}(D^\pm/D^\pm_s\to e^\pm\nu)=BR(D^\pm/D^\pm_s\to e^\pm\nu)
+BR(D^\pm/D^\pm_s\to e^\pm\nu\gamma)_{E_{\gamma}<E_{\rm res}}
\end{equation}
However, the soft photon contribution is only a small fraction of
the total rate for
BR($D^\pm/D^\pm_s\to e^\pm\nu \gamma$), and so
BR$^{\rm eff}(D^\pm/D_s^\pm\to e\nu$) would still 
be below the expected experimental sensitivity of ${\cal O}(10^{-4})$. 
Hence we suggest that observation of 
BR$(D^\pm/D^\pm_s\to e^\pm\nu)\ge 10^{-4}$
could only be attributed to physics beyond the SM.

We note that inclusive measurements of the 
BR($D^\pm_s/D^\pm \to e^\pm +X$) have
been performed, to which any enhanced $D^\pm_s/D^\pm \to e^\pm\nu$ would have
contributed. However, the error in the measurements of
$D^\pm_s\to e^\pm +X$ \cite{Bai:1997un}
and  $D^\pm\to e^\pm +X$ \cite{Abbiendi:1998ub} 
still allow for contributions from $D^\pm_s/D^\pm \to e^\pm\nu$ 
of the order of a percent or more.

%%%%%%%%%%%%%%%%%%%%%%%%%%%%%%%%%%%%%%%%%%%%%%%%%%%%%%%%%%
\boldmath
\section{$R$ parity violating contributions to $D^\pm,D_s^\pm\to \ell^\pm\nu$}
\unboldmath
%%%%%%%%%%%%%%%%%%%%%%%%%%%%%%%%%%%%%%%%%%%%%%%%%%%%%%%%%%

The main motivation for $R$ parity violating SUSY 
\cite{Dreiner:1997uz,Allanach:1999ic}
is to account for the observed neutrino oscillations without
increasing the particle content of the MSSM 
\cite{Aulakh:1982yn,Borzumati:2002bf}. The superpotential is given by:
\begin{equation} \label{potential}
W_{R}={1\over 2}\lambda_{ijk}L_iL_jE^c_k+
\lambda'_{ijk}L_iQ_jD^c_k+{1\over 2}\lambda''_{ijk}U^c_iD^c_jD^c_k
\end{equation}
Bilinear terms $\mu_iL_iH_2$ are also possible, but have 
negligible impact on the annihilation decays we consider. 
Since the $\lambda''_{ijk}U^c_iD^c_jD^c_k$ term
can mediate proton decay, it is customary to assume that 
the $\lambda''$ couplings vanish due to some discrete symmetry 
(e.g. baryon parity).
The simplest approach to $R$ parity violating phenomenology is
to assume that a single $R$ parity violating coupling
in the {\sl weak basis} $(\lambda'_{ijk}$)
is dominant with all others negligibly small. 
It was shown that such an approach leads to several non-zero 
$R$ parity violating couplings in the {\sl mass basis} 
$(\bar\lambda'_{imn})$ due to quark mixing \cite{Agashe:1995qm}:
\begin{equation}
\bar\lambda'_{imn}=\lambda'_{ijk}V^{\rm KM}_{jm}\delta_{kn}
\end{equation}
Here we have assumed that all quark mixing lies in the up--type sector,
so that the mixing matrix is the usual Kobayashi--Maskawa matrix $V^{\rm KM}$.
This simplification avoids the appearance of the right--handed quark
mixing matrix and gives the most conservative limits
on the $R$ parity violating couplings, which would otherwise
be constrained more severely from the decay $K^\pm\to \pi^\pm\nu
\overline\nu$ \cite{Agashe:1995qm}. A realistic $R$ parity violating 
model would have many non--zero couplings {\sl in the weak basis} and so 
in general would have a very rich phenomenology
provided the couplings are not too small.

\begin{figure} \begin{center}
\includegraphics[width=9cm]{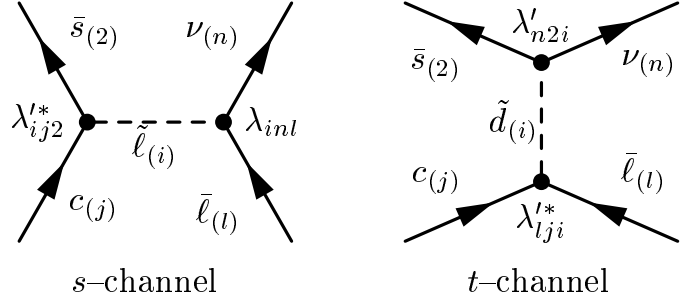}
\caption{Leptonic $D_s$ decay in $R$ parity violating models: 
sparticle exchange
\label{sparticleexchange}}
\end{center} \end{figure}

It has been emphasised before that the purely leptonic decays 
are very sensitive at tree level to 
$R$ parity violating trilinear interactions, 
and thus these decays constitute excellent probes of
the model e.g. the decays $B^\pm_{u,c}\to l^\pm\nu$  
may be enhanced up to current experimental sensitivity
\cite{Baek:1999ch,Dreiner:2001kc,Akeroyd:2002cs,Akeroyd:2002gr}.
The relevant Feynman diagrams for the decays $D^\pm/D_s^\pm\to l^\pm \nu$
are depicted in Fig.~\ref{sparticleexchange}
and consist of $s$-- and $t$--channel exchange of sparticles. 
These additional channels modify the SM rate (\ref{smrate}) by
\begin{equation}
m_l \quad\to\quad \left(1+{\cal A}^q_{ln}\right)m_l - (R+ {\cal B}^q_{ln})M_{D_q}\,,
\end{equation}
where $R_\ell=m_\ell M_{D_q} \tan^2\beta/M_{H^\pm}^2$ 
stems from $R$ parity conserving SUSY charged Higgs exchange 
\cite{Hou:1992sy} which we will not consider further
since it is also proportional to the lepton mass and is
relatively unimportant for the lighter leptons on which we focus. 
The $R$ parity violating SUSY contributions are given by:
\begin{eqnarray}
{\cal A}^q_{ln}&=&{\sqrt 2\over 4G_FV_{cq}}\sum_{i,j=1}^3{1\over 
2 m^2_{\tilde q_i}}V_{2j}\lambda'_{nqi}\lambda'^*_{lji},\\
{\cal B}^q_{ln}&=&{\sqrt 2\over 4G_FV_{cq}}\sum_{i,j=1}^3{2\over 
m^2_{\tilde \ell_i}}V_{2j}\lambda_{inl}\lambda'^*_{ijq},
\end{eqnarray} 
where $q=d,s$ for $D^+,D^+_s$, respectively. These formulae were derived in 
\cite{Baek:1999ch} for leptonic decays of $B$ mesons. 
The helicity suppressed contribution
from the ${\cal A}^q_{ln}$ term can be mediated by just one $\lambda'$
coupling (if $n=l$ and $q=j$), or by two different couplings 
(if $n\ne l$ and/or $q\ne j$).
%\footnote{We will
%discuss a special case later on} 
The dominant ${\cal B}^q_{ln}$ term
(which is not helicity suppressed) requires one non--zero $\lambda$ 
{\it and} one non--zero $\lambda'$. In the next section we will vary the 
$R$ parity violating couplings inside their allowed ranges to determine
the obtainable BRs.

The contribution of the $t$--channel diagrams has been considered
in \cite{Ledroit:1998} for $D^\pm_s\to \tau^\pm\nu$ and 
$D^\pm_s\to \mu^\pm\nu$. Here only
single coupling limits were considered and
weak limits are derived for $\lambda'_{32k}$ and $\lambda'_{22k}$. 
Analogous $t$--channel exchange diagrams also occur for the neutral 
$D^0$ meson decays $D^0\to l^+_il^-_j$ 
\cite{Burdman:1995te},\cite{Burdman:2001tf}.
Strong upper limits ($<10^{-6}$) on 
BR$(D^0\to e^+e^-,e^+\mu^-,\mu^+\mu^-$) have been obtained from 
the Tevatron. These decays have no $s$--channel contributions of the 
type $\lambda\lambda'$ due to the absence of the coupling 
$\lambda'_{ijk}\tilde \nu_i u_j\overline u_k$ in the Lagrangian.
We wish to focus on the decays $D^\pm_s,D^\pm\to e^\pm\nu$, in particular the
helicity unsuppressed $s$--channel contributions mediated by 
combinations of $\lambda\lambda'$.
Although these decays might be problematic at the Tevatron 
due to the missing energy of $\nu$, they can be readily searched for 
at the $e^+e^-$ machines BELLE, BaBar and CLEO-c.

%%%%%%%%%%%%%%%%%%%%%%%%%%%%%%%%%%%%%%%%%%%%%%%%%%%%%%%%%%
\boldmath
\section{Numerical results}
\unboldmath
%%%%%%%%%%%%%%%%%%%%%%%%%%%%%%%%%%%%%%%%%%%%%%%%%%%%%%%%%%
In our analysis we make use of the latest limits on the $R$ 
parity violating couplings $\lambda$ and $\lambda'$. Single coupling
bounds are listed in \cite{Allanach:1999ic}. Further input
parameters are $m_{D^\pm} = 1.8693\, {\rm GeV},\, \tau_{D^\pm}=1.051 
\times 10^{-12} s,\, f_{D^\pm}= 0.2\, {\rm GeV}$ for the $D^\pm$ meson
and $m_{D_s^\pm} = 1.9685\, {\rm GeV},\, \tau_{D_s^\pm}=0.49 \times 10^{-12} s,
\, f_{D_s^\pm}= 0.25\, {\rm GeV}$ for the $D_s^\pm$. The parameters
for rare $\tau$ decays are taken from \cite{Saha:2002kt}.

Many of the bounds on $R$ parity violating couplings relevant
for our analysis are of the same order of magnitude; the products 
of these couplings are ${\cal O}(10^{-2})$ and can mediate
$D^\pm/D^\pm_s\to e^\pm\nu$ with BRs up to ${\cal O}(10^{-1})$.
However, most combinations
of $\lambda\lambda'$ which mediate $D^\pm/D^\pm_s\to e^\pm\nu$
would strongly contribute to the  Kaon decays, $K^0\to e^+e^-,
e^\pm\mu^\mp,\mu^+\mu^-$, via $\tilde \nu$ exchange in the
$s$--channel \cite{Choudhury:1996ia}. The limits on such combinations 
is ${\cal O}(10^{-7})$, which is 
$10^5$ better than the product of the single coupling limits,
and at first sight would seem to rule out the possibility of 
a sizably enhanced $D^\pm/D^\pm_s\to e^\pm\nu$
mediated by $\lambda\lambda'$ combinations. 

However, large BRs for $D^\pm/D^\pm_s\to e^\pm\nu$ can occur
if the neutrino is $\nu_{\tau}$. This is because
the corresponding decay in the Kaon sector would involve
a $\tau$ lepton in the final state, which is kinematically impossible.
The only possibility for a lepton flavour violating Kaon decay mediated
by these combinations of couplings would be 
$K^0\to \ell^\pm\tau^{\mp*}\to \ell^\pm \ell'^\mp
\nu_{\tau}\nu_{\ell'}$. Here the additional suppression factors
(e.g.\ off--shell propagators, additional vertices)
easily weaken the limits on the relevant $\lambda\lambda'$
couplings to ${\cal O}(10^{-2})$ where the single coupling
bounds become more restrictive than the bounds from Kaon decays.
In addition the final state $e^\pm l^\mp \nu_{\tau}\nu_l$ has not
been searched for. 

Therefore the most promising combinations which enhance
$D^\pm\to e^\pm\nu_{\tau}$ and $D_s^\pm\to e^\pm\nu_{\tau}$
are $\lambda_{231}\lambda'_{221}$ and $\lambda_{231}\lambda'_{222}$
respectively. 
For these combinations the use of the single coupling 
bounds is justified.\footnote{Recently some new constraints on 
these combinations
have been derived from considering 2 loop contributions to 
neutrino masses \cite{Borzumati:2002bf}. These bounds are of the same
order of magnitude as the single coupling bounds, and so our results 
are largely unaffected.} The single coupling
bounds $\lambda_{231}= 0.07$ and $\lambda'_{221}= 0.18$  (for
sparticle mass $100$ GeV) can induce
BR$(D^\pm\to e^\pm\nu_{\tau})= 1.251 \cdot 10^{-2}$. Although
as stated above, this combination of couplings is safe from rare Kaon
decay bounds, the same combination can induce the lepton flavour
violating $\tau$ decay $\tau^\pm\to e^\pm K^0_s$ \cite{Saha:2002kt}.

\begin{figure}
\begin{center}
\includegraphics[width=15cm]{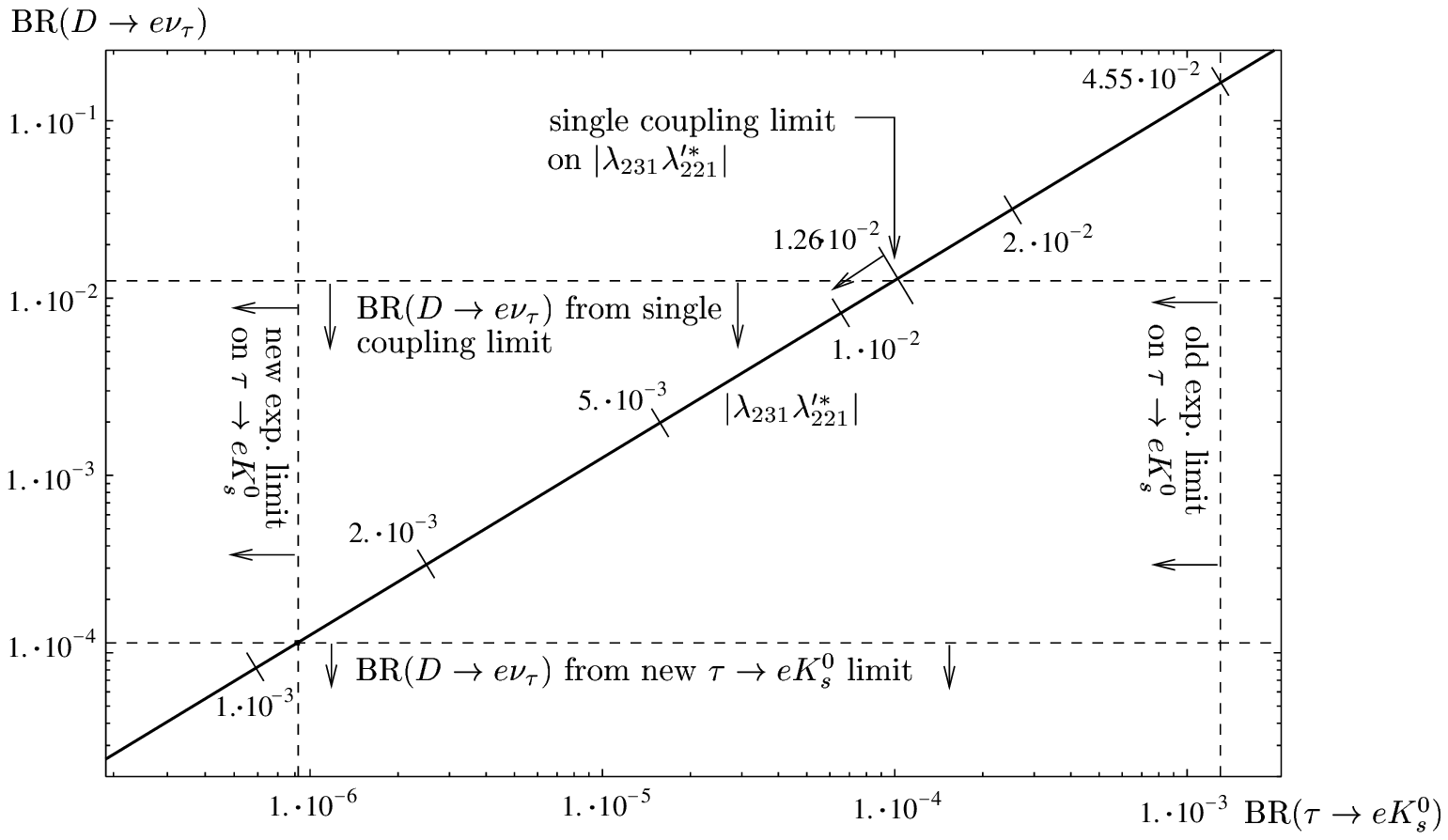}
\end{center}
\caption{Dependence of BR($D^\pm\to e^\pm\nu_\tau$) and BR($\tau^\pm\to e^\pm K^0_s$)
on the product of $\protect\fmslash R_p$ couplings $\left|\lambda_{231}\lambda'^*_{221}\right|$.
\label{Ddcorrelation}}
\end{figure}

The dependence of BR($D^\pm\to e^\pm\nu_\tau$) and 
BR($\tau^\pm\to e^\pm K^0_s$)
on the product of the $\fmslash R_p$ couplings 
$\left|\lambda_{231}\lambda'^*_{221}\right|$
is shown in Fig.\ref{Ddcorrelation}. 
The $x$-- and $y$--axes give the respective branching ratios and 
$\left|\lambda_{231}\lambda'^*_{221}\right|$ is varied along the diagonal line.
The plot is logarithmic along both axes.
The experimental bound on BR($\tau^-\to e^-K_0$) prior to summer 2002 
\cite{Hayes:1981bn}
(rightmost vertical dashed line) is less restrictive for 
$\left|\lambda_{231}\lambda'^*_{221}\right|$, and therefore 
for BR($D^\pm\to e^\pm\nu_\tau$),
than the product of the individual coupling bounds. The BR($D^\pm\to e^\pm\nu_\tau$)
attainable with the individual coupling bounds is indicated by the upper horizontal
dashed line. However, a much improved experimental bound on
BR($\tau^\pm\to e^\pm K^0_s$) has recently been published 
\cite{Chen:2002ug}, and this limit is
indicated by the left vertical dashed line.\footnote{\cite{Saha:2002kt}
appeared shortly before the new bounds on BR($\tau^\pm\to e^\pm K^0_s$)
were released and therefore derives a rather weak bound
$\left|\lambda_{231}\lambda'^*_{212}\right|,
\left|\lambda_{231}\lambda'^*_{221}\right| < 4.7 \times 10^{-2}$, 
which is less
restrictive than the single coupling limit. This limit improves to
$1.2 \times 10^{-3}$, an order of magnitude better than the single
coupling limit, with the new data from \cite{Chen:2002ug}.}
This new limit restricts the enhancement of
BR($D^\pm\to e^\pm\nu_\tau$) from $\fmslash R_p$ couplings quite 
substantially;
only BRs of ${\cal O}(10^{-4})$ remain attainable, while the older bound
on lepton flavour violating $\tau$--decays allowed BRs of ${\cal O}(1\%)$.
CLEO-c expects 30 million $D^\pm$ events (6 million tagged), so even BRs of
${\cal O}(10^{-4})$ or smaller could be observed. Alternatively, lack of
observation would further improve the limit on $\left|\lambda_{231}\lambda'^*_{221}\right|$.

\begin{figure}
\begin{center}
\includegraphics[width=15cm]{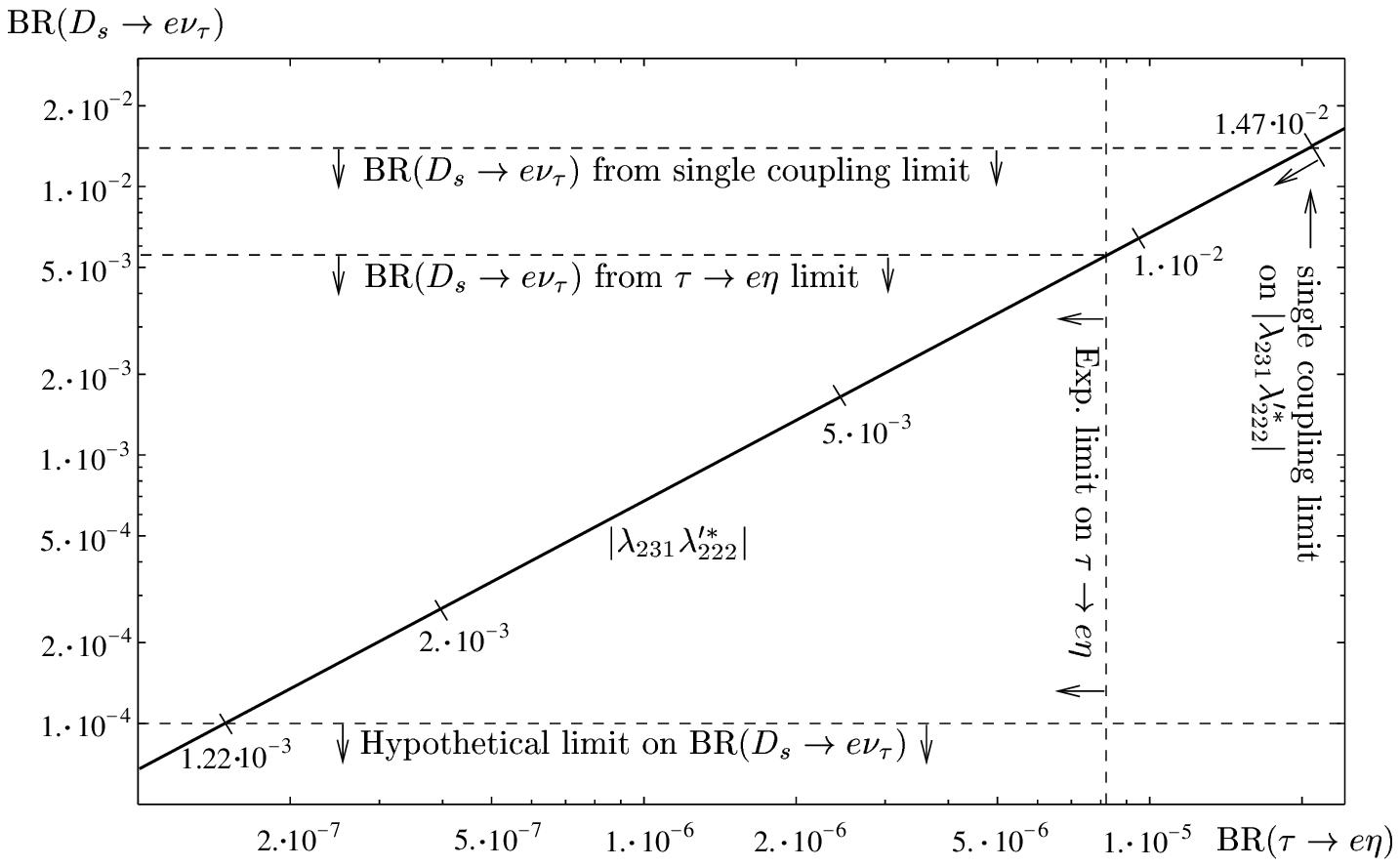}
\end{center}
\caption{Dependence of BR($D_s\to e\nu_\tau$) and BR($\tau\to e\eta$)
on the product of $\protect\fmslash R_p$ couplings $\left|\lambda_{231}\lambda'^*_{222}\right|$.
\label{Dscorrelation}}
\end{figure}

The situation is much more favourable for $D_s^\pm\to e^\pm\nu_\tau$,
which is correlated with the less well measured $\tau^\pm\to e^\pm\eta$.
The single coupling bounds 
$\lambda_{231}= 0.07$ and $\lambda'_{222}= 0.21$
give BR$(D_s^\pm\to e^\pm\nu_{\tau})= 1.391 \cdot 10^{-2}$.
The plot analogous to Fig.\ref{Ddcorrelation} 
for the dependence of BR($D_s^\pm\to e^\pm\nu_\tau$) and 
BR($\tau^\pm\to e^\pm\eta$) on the product of $\fmslash R_p$ couplings 
$\left|\lambda_{231}\lambda'^*_{222}\right|$ is shown in 
Fig.\ref{Dscorrelation}.
The current experimental bound on BR($\tau^\pm\to e^\pm\eta$) 
\cite{Bonvicini:1997bw}
(vertical dashed line) restricts 
$\left|\lambda_{231}\lambda'^*_{222}\right|$ (and therefore BR($D_s^\pm\to e^\pm\nu_\tau$))
slightly more than the product of the single coupling limits, indicated in the
top right hand corner of the graph.

The lower horizontal dashed line indicates a hypothetical 
limit on BR($D_s^\pm\to e^\pm\nu_\tau$) of $10^{-4}$ which seems realistic
in light of an expected number of 1.5 million $D_s$ events (0.3 million fully
tagged) after one year of running of CLEO-c. 
This limit would restrict the product of couplings 
$\left|\lambda_{231}\lambda'^*_{222}\right|$ about a factor of 10 better than
present experiments. To compete with this accuracy, the experimental limit
on BR($\tau^\pm\to e^\pm\eta$) would have to improve by about two orders of magnitude
which does not seem attainable in the current runs of the $B$ factories.
We therefore believe that even if searches for 
$D^\pm/D_s^\pm\to e^\pm\nu_\tau$ do not
detect an enhancement in these channels, they would still be useful for setting
new limits on products of $\fmslash R_p$ couplings.

The poorly measured decay $D^\pm\to \mu^\pm\nu$ may also be
enhanced by the combination $\lambda_{232}\lambda'_{221}$.
The current error allows for a sizeable enhancement over
the SM prediction of the order 3--6, depending on the value
of the decay constant. 
The SM prediction currently lies at the lower end of the
experimentally allowed interval. If subsequent measurements
should tend towards the upper end of the current interval,
non--zero $R$ parity violating couplings would be a possible
explanation for this deviation. 
SM--conform measurements on the other hand would allow for
better limits on $\lambda_{232}\lambda'_{221}$.
CLEO-c aims to measure the
BR($D^\pm\to \mu^\pm\nu$) to a precision of a few $\%$.  
For the decays $D^\pm/D_s^\pm\to \tau^\pm\nu$ we do not find
sizably enhanced BRs.
 
The decays $D^\pm\to \ell^\pm\nu$ and $D_s^\pm\to \ell^\pm\nu$ can also
be mediated by products of two $\lambda'$ couplings (right diagram in
Fig.\ref{sparticleexchange}), but because of the helicity suppression,
only the $\tau$--channel can receive a sizeable contribution. Even in
these cases, the $\fmslash R_p$ contributions can only become as large
as the uncertainties of the SM predictions. Therefore neither a large 
enhancement, nor improvements of the limits are possible, except 
for single coupling limits on $\lambda_{22k}'$ and 
$\lambda_{32k}'$ as shown in \cite{Ledroit:1998}.

\section{Conclusions}
In the context of the MSSM we have studied the effects of 
$R$ parity violating couplings ($\lambda,\lambda'$)
on the purely leptonic decays $D^\pm/D_s^\pm\to l^\pm\nu$.
We showed that slepton mediated contributions proportional
to combinations of the type $\lambda\lambda'$ can strongly
enhance the previously unmeasured decays $D^\pm/D_s^\pm\to e^\pm\nu$
to the sensitivity of current $B$ factories and forthcoming CLEO-c.
Maximum values for BR($D^\pm_s\to e^\pm\nu$) and BR($D^\pm\to e^\pm\nu$) of 
$5\times 10^{-3}$ and $1\times 10^{-4}$ respectively 
were found. Mere observation of these
decays would be an unequivocal signal of physics beyond the SM.
In simple $R$ parity violating models with a single dominant
$\lambda\lambda'$ combination, there would be a correlation with
the decays $\tau^\pm\to e^\pm K^0_S$ and $\tau^\pm\to e^\pm\eta$, which
would be similarly enhanced to the sensitivity of current and 
planned experiments. Such correlated
signals would provide strong evidence for $R$ parity violating interactions.

\vspace{20mm}
\begin{center}
{\large\bf  Acknowledgements} 
\end{center}

The authors wish to thank F.~Borzumati, S.~Pakvasa and B.~Yabsley, 
for useful discussions and comments.
S.R.\ was supported by the Japan Society for the Promotion
of Science (JSPS).

\renewcommand{\theequation}{B.\arabic{equation}}
\setcounter{equation}{0}

\end{document}